\documentclass[prb,showpacs,superscriptaddress,titlepage,amsmath,amssymb,twocolumn,floatfix]{revtex4-1}
\usepackage{caption}
\usepackage{graphicx} 
\usepackage{subfigure}  
\usepackage{dcolumn}   
\usepackage{bm}  
\usepackage{epstopdf} 
\usepackage{bbold}
\usepackage{slashed}
\usepackage{array}
\usepackage{booktabs}
\begin{document}
\title{A Wilsonian Approach to Crystal Structure Transformations driven by Strong Electron Correlations}
\author{J.~M.~Booth}
\email{jamie.booth@rmit.edu.au}
\affiliation{ARC Centre of Excellence in Exciton Science, RMIT University, Melbourne, Australia}
\affiliation{Theoretical Chemical and Quantum Physics, RMIT University, Melbourne, Australia}

\begin{abstract}
	Mathematical descriptions of the interplay between strong electron correlations and lattice degrees of freedom are of enormous importance in the development of new devices based on metal oxides such as VO$_{2}$ and the Cuprate superconductors. In this work the physics of tight-binding type electron momentum states interacting with lattice fluctuations is reformulated into an approach based on lattice QCD. Strong electron correlations act as a source for phonons, which are incorporated by using SU(2) bosons acting on neighboring atomic sites. This allows the system to be described by a Hamiltonian which describes strong interactions between SU(2) Yang-Mills bosons near T$_{c}$ resulting from electron correlations. Monte Carlo and GW calculations show that at low Temperature the electron-electron interactions drive the system into a phase coherent phonon state, breaking the lattice symmetry, and a band gap opens. This formalism is intrinsically able to combine strong-electron correlations with lattice fluctuations in a manner which describes symmetry-breaking structural phase transitions which manifest spin- and charge ordering.
\end{abstract}

\maketitle
\section{Introduction}
Over the course of the last century tremendous progress has been made in describing the physics of materials using first quantum mechanics, and subsequently ideas and machinery from Quantum Field Field Theory. However, despite this progress, and the huge effort which has been poured into this search for understanding, precise descriptions of certain phenomena have remained stubbornly intractable. 

One such problem is a mathematical description of structural phase transitions which occur in systems in which strong electron correlations are also present. Such systems exhibit metal-insulator transitions involving spin and charge ordering\cite{Imada1998} which have huge potential for new generations of electronic devices.\cite{Liu2018,Shao2018}

The metal-insulator-structural phase transition of vanadium dioxide is one such system. VO$_{2}$ is a 3$d^{1}$ system which exists in a metallic, tetragonal structure above $\sim$ 340 K, and when pure or unstrained adopts a monoclinic P2$_{1}$/c structure (usually called ``M$_{1}$") below T$_{c}$.\cite{Goodenough1971} 

There has been an intense debate about the nature of the insulating phase raging for some decades, as the atomic rearrangements in going from Tetragonal to the Monoclinic structure are highly reminiscent of the Peierls mechanism: the vanadium chains which run parallel to the tetragonal c-axis dimerize (and also experience an antiferroelectric distortion,\cite{Pouget1974} see Figure 1a-b), which suggest a band description of the metal-insulator transition from perturbation theory.\cite{Goodenough1971} However the tetragonal electron liquid exhibits the characteristics of a strongly correlated system near T$_{c}$\cite{Qazilbash2007} and thus the insulating state resulting from this would be expected to be a Mott-Hubbard insulator.\cite{Hubbard1963} 

Strong support for the Mott scenario was voiced by Mott himself\cite{Mott1974} due to the appearance of the ``M$_{2}$" structure upon doping with holes, extra electrons or by inputting stress/strain.\cite{Pouget1974,Park2013} This structure is also Monoclinic (although C2/m rather than P2$_{1}$/c),\cite{Pouget1974} however in this form the dimerization and antiferroelectric distortion occur individually, and not together, each occurring on a neighbouring vanadium chain (see Figure 1c). 

There is no doubt however that the M$_{2}$ form \textit{is} a Mott insulator, as DMFT\cite{Brito2016} and modified GW calculations\cite{Booth2016a} confirm, however the most significant aspect of this resolution is that the antiferroelectric chains are also antiferromagetically ordered, while the dimerized chains show no such order.\cite{Pouget1976} 

This coincidence of antiferroelectricity and antiferromagnetism led the author to propose a Yang-Mills description of electron-phonon interactions in metal oxides\cite{Booth2020}, which can manifest spin-ordering via the transverse phonons, and pairing via the longitudinal phonons separately. In that study the basic interaction structure was determined, including the spinor grouping, however the adaptation to mass generation, i.e. metal-insulator transitions, focused on the electronic structure, and ignored the underlying lattice.

\begin{figure}[h!]
	\subfigure{\includegraphics[width=0.5\columnwidth]{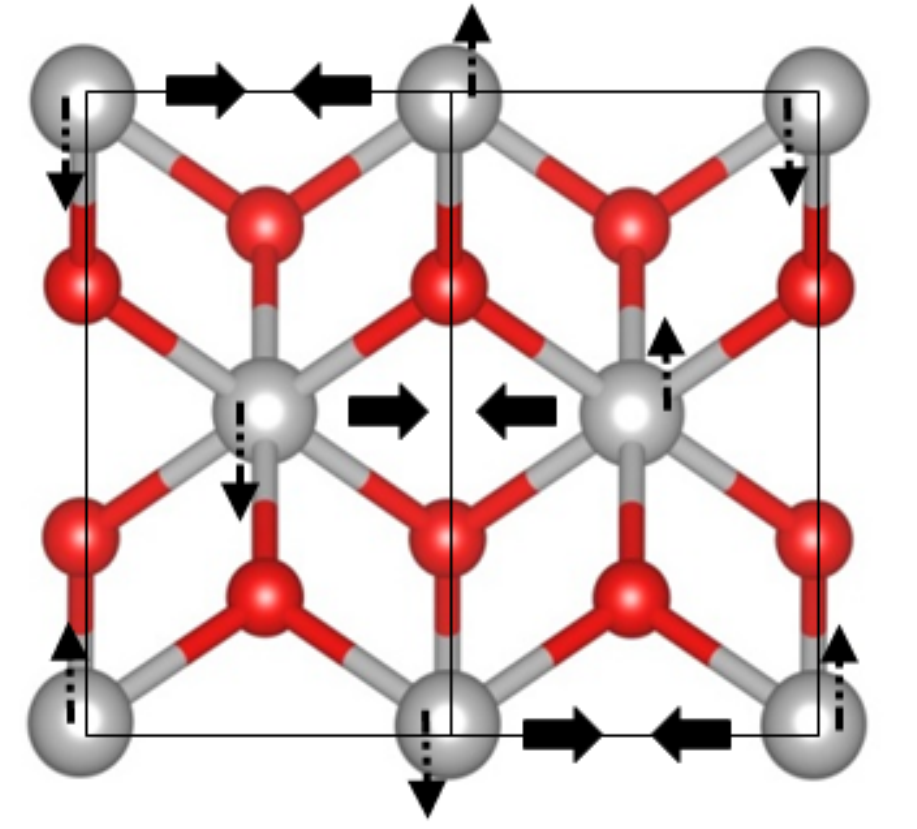}}{a)}\\
	\subfigure{\includegraphics[width=0.5\columnwidth]{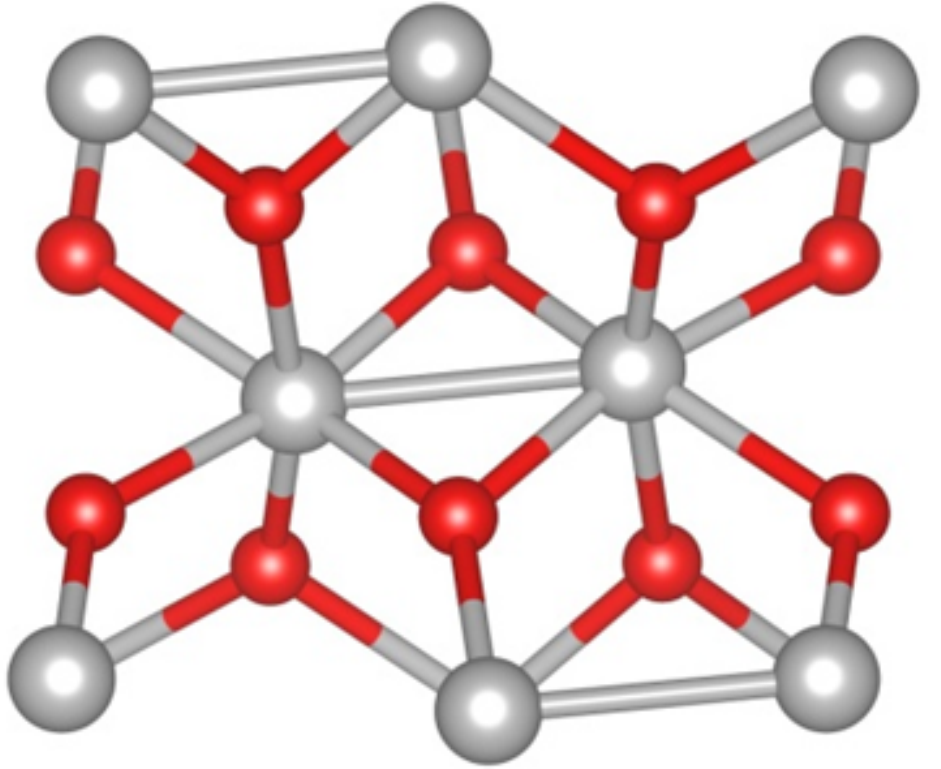}}{b)}\\
	\subfigure{\includegraphics[width=0.8\columnwidth]{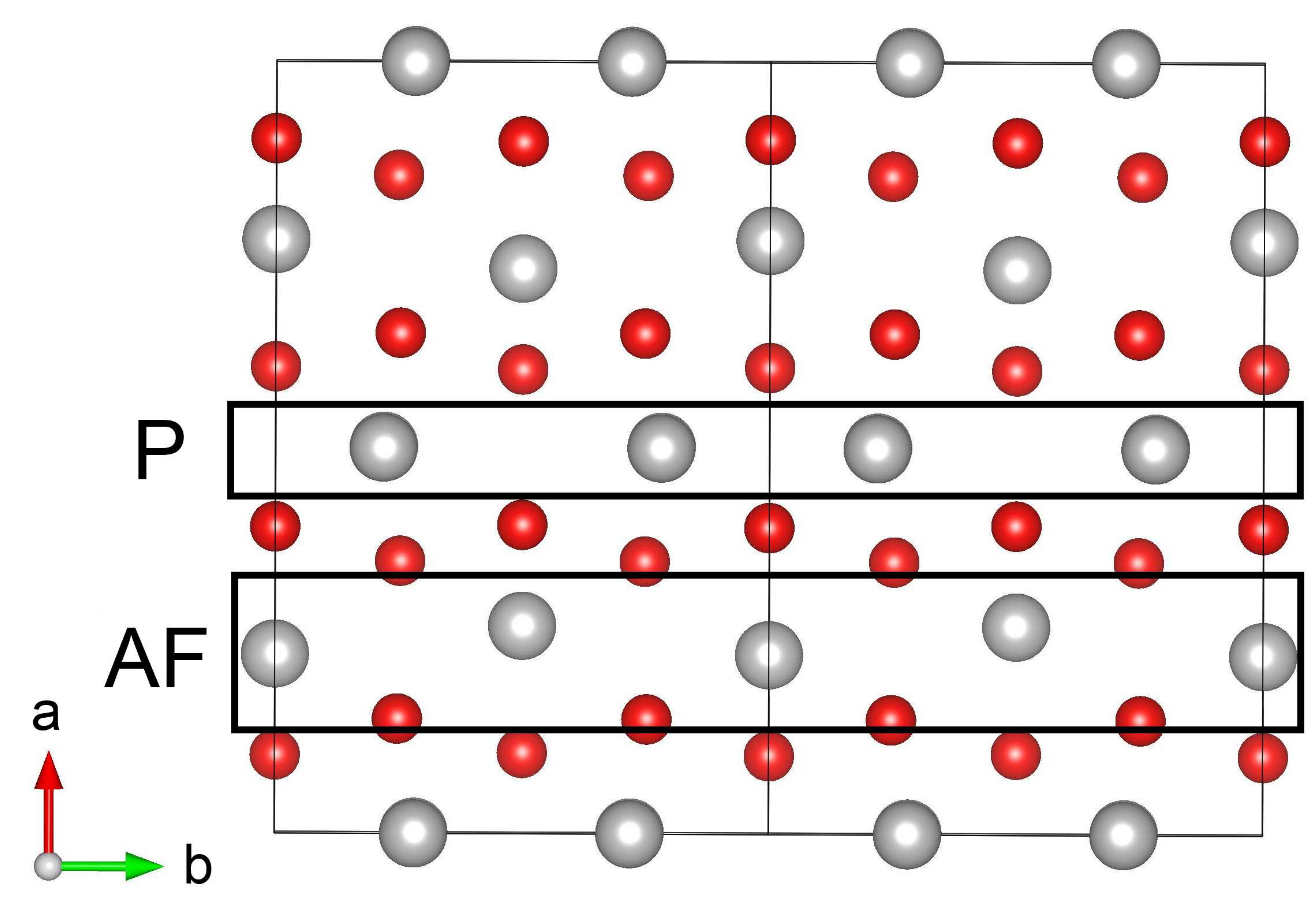}}{c)}
	\caption{\raggedright{a) Crystal structure of the tetragonal form of vanadium dioxide with the distortions which create the M$_{1}$ monoclinic structure indicated by arrows, b) resultant M$_{1}$ structure, and c) crystal structure of the M$_{2}$ from of VO$_{2}$ with the Peierls paired and antiferroelectrically distorted chains indicted by ``P" and ``AF" respectively.}}
	\label{M2}
\end{figure}
While the use of Yang-Mills theory may seem at first glance an unnecessary complication, microscopic mechanisms of crystal structure transformations which are accompanied by spin ordering are thin on the ground. The Landau approach\cite{Landau2008} of defining an order parameter which the Free Energy of the system is a function of (often a simple, even polynomial) is a crude abstraction which may hide similarities between systems in degrees of freedom which the approach integrates out.

However, it is obvious that a more complete description of the electron-phonon mechanism which describes spin and charge ordering along with metal-insulator transitions cannot be a simple Yukawa potential: $V = -g\phi\bar{\psi}\psi$,\cite{Zee_Yukawa} in which a scalar field interacts with the electrons. The atomic displacements characterizing the VO$_{2}$ system for example are vectors, and the scalar boson $\phi$ has no structure which can act on the spinor variables. In addition this potential does not contain a description of the symmetry-broken state.

To successfully describe a transition from a paramagnetic and metallic, high symmetry state to an insulating low symmetry state with a specific spin ordering (singlets or antiferromagnetism for example) a theory needs:

\textit{i)} Vector bosons. The transition from one ordered crystal structure to another will involve rearrangements of the atoms which will be described by vectors. These vector fluctuations gaining a vacuum expectation value will generate the symmetry-broken ground state.

\textit{ii)} Pauli- or Gamma matrices which couple the vector bosons to the electron spinor variables such that spins can be ordered. 

\textit{iii)} An interaction vertex between the spinors and the vector bosons which can act on the high symmetry state and produce the correct charge and spin ordering of the low symmetry state. This will be a function of the dominant interactions in the system. For example, in this work, this vertex arises due to strong electron correlations, and is matrix valued, which provides the correct symmetry-breaking form.

In the following discussion a formalism to describe crystal lattices and their fluctuations, including crystal structure transformations which are accompanied by spin ordering and metal-insulator transitions is developed based on an SU(2) Yang-Mills theory. This theory is adapted to a non-relativistic lattice environment by first reformulating the mathematical description of electronic states interacting with lattice fluctuations into an approach which is an adaptation of lattice gauge theory.

We find that a remarkable coincidence exists between the interaction vertex structure described above and the SU(2) theory used to describe the weak interaction of the Standard Model of particle physics.\cite{Srednicki2007} If the SU(2) generators are re-imagined as 2$\times$2 linear transformations acting on neighboring sites of a quasi-linear system the interaction vertex has all of the properties required. 

We use the example of vanadium dioxide to motivate the development, however the formalism applies to any quasi-linear system in which the electron behavior is governed by a Hubbard Hamiltonian. An SU(3) description of hexagonal systems such as graphene is presented in a different study.\cite{Booth_SU3}

\section{Results and Discussion}
\subsection{Spinor Action}
\subsubsection{Derivatives}
The starting point for developing a new formalism is the high symmetry state, and the recognition that as a crystal system it has well-defined momentum states. In the tight-binding representation a momentum state is given by:
\begin{equation}
\psi_{\mathbf{k}}(\mathbf{r})=\sum_{\mathbf{R}}\phi(\mathbf{r}-\mathbf{R})e^{i\mathbf{kR}}
\label{momentum_state}
\end{equation}
where $\phi(\mathbf{r})$ is a single or a sum of position state wavefunctions such as orbitals, and $\mathbf{R}=n\mathbf{a}$ is a lattice vector, i.e. some integer multiple of the unit cell constant. We see that in this representation, the lattice enters in the phase which is applied to the position state wavefunctions.
To parallel transport we need a Unitary: $U(x_{ij},\mu)$, which translates our wavefunction, from which we can construct a derivative:
\begin{equation}
i\partial_{\mu}\psi(x_{i})=\frac{1}{|x_{j}-x_{i}|}\big(U(x_{ij},\mu)\psi(x_{j})-\psi(x_{i})\big)
\end{equation}
where $x_{i},x_{j}$ are neighbouring nuclei position vectors and their difference ($x_{j}-x_{i}$) is a unit cell  vector, and the definition of the momentum state (equation \ref{momentum_state}) requires only the forward derivative, and thus no Fermion doubling problem arises.\cite{Gattringer2013_1}

From the definition of our momentum state we have that:
\begin{equation}
U(x_{ij},\mu)\psi(x_{j}) = e^{ik^{\mu}(\hat{x}_{j,\mu}-x_{i,\mu})}\hat{T}(x_{j}-x_{i})\psi(x_{j})
\end{equation}
where $\hat{T}$ is the translation operator, and the derivative becomes:
\begin{multline}
i\psi^{*}(x_{i})\partial_{\mu}\psi(x_{i})=\frac{1}{|x_{j}-x_{i}|}\times\\\langle\psi(x_{i})|\mathbb{1}-e^{ik^{\mu}(x_{j,\mu}-x_{i,\mu})}\hat{T}(x_{j}-x_{i})|\psi(x_{j}\rangle
\label{Spinor_Deriv}
\end{multline}
This formulation assumes that the dominant contribution to the momentum state comes from the nearest position state wavefunction, which for tight-binding-type wavefunctions and the purposes of this work will be sufficient.

\subsubsection{Lattice Fluctuations and Scattering}
This formulation allows us to introduce fluctuations in the positions of the atoms rather easily:
\begin{multline}
U(x_{ij},\mu) = e^{ik^{\mu}((x_{j,\mu}+\delta x_{j,\mu})-(x_{i,\mu}+\delta x_{i,\mu}))}\times\\\hat{T}((x_{j}+\delta x_{j})-(x_{i}+\delta x_{i}))
\label{Unitary_Fluct}
\end{multline}
The fluctuations will be small (or the lattice will be destroyed) and thus we can write:
\begin{multline}
e^{ik^{\mu}((x_{j,\mu}+\delta x_{j,\mu})-(x_{i,\mu}+\delta x_{i,\mu}))}\rightarrow \\ e^{ik^{\mu}(x_{j,\mu}-x_{i,\mu})}(1+ik^{\mu}\delta x_{j,\mu})(1-ik^{\mu}\delta x_{i,\mu}) \\
= e^{ik^{\mu}(x_{j,\mu}-x_{i,\mu})}(1+ik^{\mu}\delta x_{j,\mu}-ik^{\mu}\delta x_{i,\mu} + k^{2}\delta x_{j}\delta x_{i})
\label{smallgexp}
\end{multline}
Where the translation operator is implied to act only on derivative terms and is therefore omitted for clarity, as for scattering terms we will be overlapping with different position states.
Since the coupling of phonons to electron momentum states is highly wavevector dependent, we define a coupling parameter $g(k)$ which contains the momentum and the constant phase term:
\begin{equation}
ie^{ik^{\mu}(x_{j,\mu}-x_{i,\mu})}k\delta x_{j,\mu} \rightarrow ig(k)\delta x_{j,\mu}
\end{equation}

Assuming that the fluctuations are local we can proceed by quantizing the displacement in the usual way, we shift $\delta x_{i,\mu}\rightarrow \hat{W}_{\mu}(x_{i})$ where:
\begin{equation}
\hat{W}_{\mu}(x) = \int \frac{d^{3}\mathbf{p}}{{2\pi}^\frac{3}{2}2E_{\mathbf{p}}^{\frac{1}{2}}}\big[\hat{a}_{\mathbf{p}}\epsilon_{\mu}(p)e^{ipx} + \hat{a}_{\mathbf{p}}^{\dagger}\epsilon^{*}_{\mu}(p)e^{-ipx}\big]
\label{A-field}
\end{equation}
and we use 4-vector notation for the polarization vector (i.e. the phonon eigenvector) $\epsilon_{\mu}(p)$ to keep the notation consistent with the derivative term, however throughout this work $\epsilon_{0}(p) = 0$ for all $p$.

Now the $\hat{W}_{\mu}(x)$ field acts on the positions of the nuclei to which $\psi(x_{i})$ and $\psi(x_{j})$ are bound. Ignoring the quadratic term and acting with this on the $\psi(x_{i}),\psi(x_{j})$, we get the original derivative, plus some new terms: 
\begin{multline}
i\partial_{\mu}\psi(x_{i}) + ig(k)\hat{W}^{*}_{\mu}(x_{i})\psi(x_{i}) + ig(k)\hat{W}_{\mu}(x_{j})\psi(x_{j})
\end{multline}
where now $\hat{a}^{\dagger}/\hat{a}$ create and annihilate polarization vectors $\epsilon^{\lambda}_{\mu}(p)$ which describe the motion of the atomic position, and the $\lambda$ are the basis vectors of the spacetime. The variation in space and time is given by the $e^{-ip_{\mu}x^{\mu}}$ term. 

However, $\psi(x)$ is a spinor, and thus the derivative $\partial_{\mu}\psi(x)$ needs to give us information on the change of the spinor variables as a function of space and time. This requires resolving the spinor variables into 4-components, a time component, and three spatial components, such that their changes can be evaluated. 

The spatial components are just the spin vector $\mathbf{S}$ which are resolved using the Pauli matrices, and the time component is just given by the identity matrix. Thus the derivative term becomes $\sigma^{\mu}\partial_{\mu}$, and accounting for helicity (i.e. hole terms) means that we double stack the Pauli matrices into gamma matrices (and change the signs on the spatial matrices of the left-handed spinor to make sure the Weyl equation is still obeyed), and now the bosons act on the Nambu spinors. For a detailed exposition of this see Booth and Russo\cite{Booth2020} however for the purposes of this work these details are unnecessary, as we work around the spinor action by incorporating it into the boson action in the form of SU(2) bosons. The derivative becomes:
\begin{equation}
i\psi^{*}(x_{i})\partial_{\mu}\psi(x_{i}) \rightarrow i\bar{\psi}(x_{i})\gamma^{\mu}\partial_{\mu}\psi(x_{i})
\end{equation}
and the gamma matrices are expressed (in the chiral basis) in two-component form as:
\begin{equation}
	\gamma^{0} = \begin{pmatrix}0&\mathbb{1}\\\mathbb{1}&0\end{pmatrix}, \gamma^{i} = \begin{pmatrix}0&\sigma^{i}\\-\sigma^{i}&0\end{pmatrix}
\end{equation}
and $\bar{\psi} = \psi^{\dagger}\gamma^{0}$.

However we now also have terms in which the fluctuating spinor position wavefunctions are overlapped with other sites, giving scattering terms:
\begin{multline}
ig(k)\bar{\psi}(x_{j})\gamma^{\mu}\hat{W}^{*}_{\mu}(x_{i})\psi(x_{i}) + \\ ig(k)\bar{\psi}(x_{i})\gamma^{\mu}\hat{W}_{\mu}(x_{j})\psi(x_{j})
\end{multline}
Grouping these terms into operators which act on the same sites, the spinor contribution to the lagrangian at each site is then:
\begin{equation}
\mathcal{L}_{spinor} = i\bar{\psi}\gamma^{\mu}(\partial_{\mu}+g(k)\hat{W}_{\mu})\psi
\end{equation}
where the position labels are dropped, and it is implied that the derivative- and scattering terms are overlapped with the same site, and neighbouring sites respectively. The action is as usual:
\begin{equation}
S_{spinor} = \sum_{x,t}  \mathcal{L}_{spinor}
\end{equation}
Thus this lattice formulation is now equivalent to the usual covariant derivative of an interacting set of fermion and vector boson fields.

\begin{figure}
\includegraphics[width=\columnwidth]{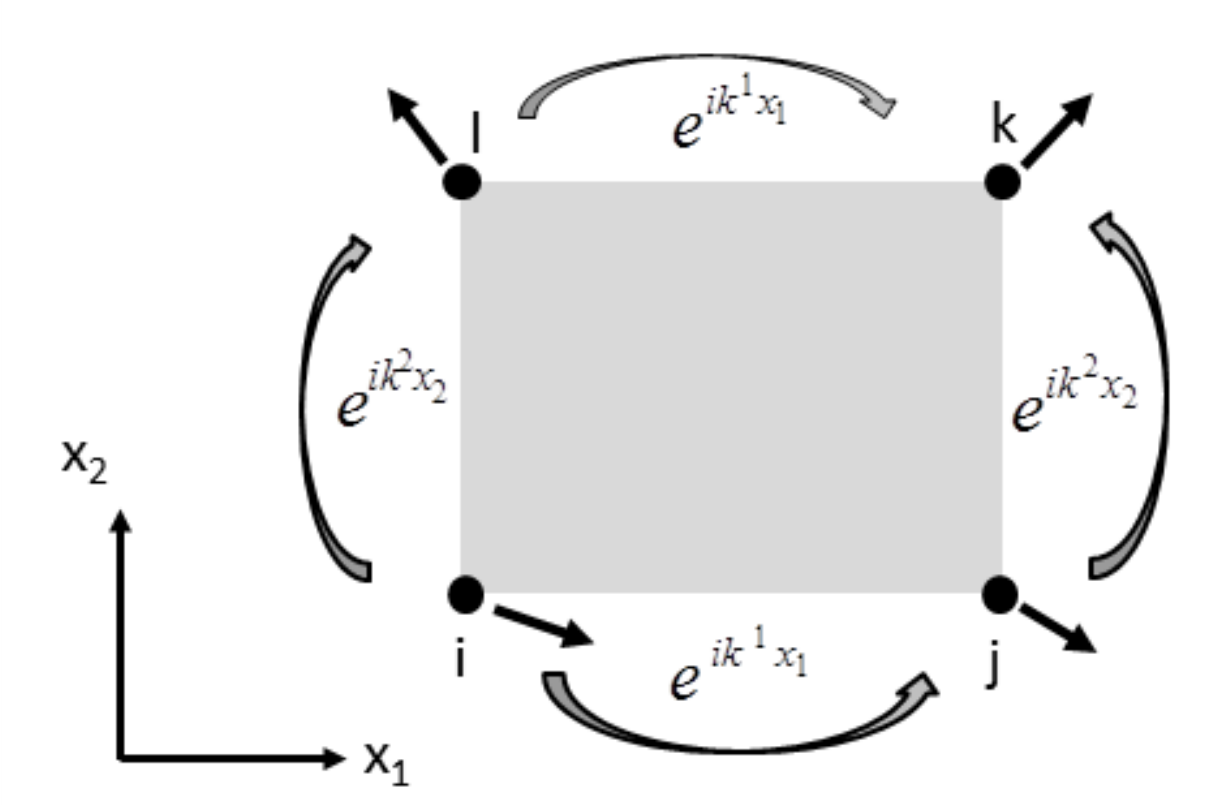}
\caption{\raggedright{Pictorial illustration of the plaquette form of the U(1) boson action. As the plaquette is traversed the exponential terms describe the changes in the polarization vector components.}}
\label{Plaq}
\end{figure}

\subsection{Boson Action}
An action for the bosonic sector can be constructed using a modification of the usual Wilsonian Lattice approach by defining plaquettes, and summing over the squares of the parallel transporters evaluated around them. While computationally a plaquette-based approach is inefficient, in this work it is used to bring out the symmetry-breaking caused by the electron correlations more clearly. 

The plaquette formulation starts from the usual definition of the parallel-transporting Unitaries defining transport around a plaquette, $x \rightarrow x+\hat{x}_{\mu} \rightarrow x+\hat{x}_{\mu}+\hat{x}_{\nu} \rightarrow x+\hat{x}_{\nu} \rightarrow x$, in a similar manner as lattice QCD.\cite{Zee_Lattice_Gauge}

However while it might seem obvious to proceed in exactly the same manner as lattice QCD for phonons using $e^{ik\hat{A}_{\mu}(x)}$ instead of $e^{\hat{A}_{\mu}(x)}$ (see Supporting information), the boson field strength $F_{\mu\nu}$ is antisymmetric for relativistic field theories. This has the consequence that massless vector fields do not contain longitudinal components, only massive vector fields do, and the longitudinal components are then supplied by the Higgs upon symmetry-breaking. However, acoustic longitudinal phonons are an experimental fact, and this requires us to proceed in a similar manner to how the spinor sector of the Lagrangian was formulated.

From equation \ref{A-field} we see that the structure of the boson momentum states is similar to those of the electron states. On a discrete lattice the field contains terms:
\begin{equation}
\hat{A}(p)\sim \sum_{x}\epsilon_{p}(x)e^{ipx}
\end{equation}
which are a prescription for building the state which is: take a polarization vector, increment one lattice spacing and multiply by a phase. i.e. if sites $i$ and $j$ are neighbors:
\begin{equation}
\epsilon_{p}(x_{j}) = \epsilon_{p}(x_{i})e^{ip(x_{j}-x_{i})}
\end{equation}
Therefore the change in the polarization vector is given by the exponential terms. Thus our Unitary is again an exponential: $\hat{U} = e^{ikx} = e^{i(\omega t - \mathbf{k}\mathbf{x})}$, although we explicitly include time dependence, as this gives the kinetic energy of the mode. Therefore the action for the bosonic sector uses the polarization vector instead of the spinor wavefunction, and computes its changes as functions of space and time.

Figure \ref{Plaq} gives a pictorial representation of the plaquette action of the lattice. Summing the changes in the polarization vectors around each plaquette gives an estimate of the curvature of the lattice, and the energy of the system. The plaquette term is:
\begin{multline}
O_{ijkl} = \sum_{k^{1}}\big(1- e^{ik^{1}x_{1}}\big)^{2}\big\rvert_{i\rightarrow j} + \sum_{k^{2}}\big(1- e^{ik^{2}x_{2}}\big)^{2}\big\rvert_{j\rightarrow k}\\ + \sum_{k^{1}}\big(1 - e^{ik^{1}x_{1}}\big)^{2}\big\rvert_{l\rightarrow k} + \sum_{k^{2}}\big(1 - e^{ik^{2}x_{2}}\big)^{2}\big\rvert_{i\rightarrow l}
\end{multline}
where $k^{i}$ is a component of the 4-vector $(\omega,k^{x},k^{y},k^{z})$. This is similar to lattice QCD if we recognise that when traversing around a loop we need $\hat{U}^{-1}$ for the segments $k \rightarrow l$ and $l \rightarrow i$, i.e. the derivatives go the opposite way to the path around the loop for those two sections. 

This plaquette term is equivalent to summing the squares of the derivatives with respect to time and space, as the $x_{i}$ are unit vectors in the respective directions, i.e. for $x_{i} = t = x_{0}$:
\begin{multline}
\lim\limits_{k\rightarrow 0}\frac{1}{x_{0}^{2}}\big(1- e^{ik^{0}x_{0}}\big)^{2} = \frac{1}{x_{0}^{2}}\big(1-(1+ik^{0}x_{0})\big)^{2}\\ = -(\eta^{00}k_{0}k_{0}) = -\bigg(\frac{\partial}{\partial \omega}\bigg)^{2}
\end{multline}
while for $x_{i} = x$:
\begin{multline}
\lim\limits_{k\rightarrow 0}\frac{1}{x_{1}^{2}}\big(1- e^{ik^{1}x_{1}}\big)^{2} = \frac{1}{x_{1}^{2}}\big(1-(1+ik^{1}x_{1})\big)^{2} \\ = -(\eta^{11}k_{1}k_{1}) = \bigg(\frac{\partial}{\partial x}\bigg)^{2}
\end{multline}
and we have:
\begin{equation}
-\bigg(\bigg(\frac{\partial}{\partial \omega}\bigg)^{2}-\bigg(\frac{\partial}{\partial x}\bigg)^{2}-\bigg(\frac{\partial}{\partial y}\bigg)^{2}-\bigg(\frac{\partial}{\partial z}\bigg)^{2}\bigg)
\end{equation}
where the coefficients $\eta^{\mu\nu}$ are given by the Minkowski metric:  
\begin{equation}
\eta^{\mu\nu} = \begin{pmatrix}
1&&&\\&-1&&\\&&-1&\\&&&-1
\end{pmatrix}
\label{eta}
\end{equation}
This gives the difference between the kinetic and potential energies, as required.

For a mode with $k = 0$ the exponential terms are all unity and thus the plaquette returns an energy of zero for translation of the coordinate system, as it should. However this finite difference operator also works for high $k$. In the case of a zone edge mode for example: $\mathbf{k} = k^{x} = \pi/a$ and setting $a = 1$ because the $x_{i}$ are unit vectors:
\begin{equation}
1-e^{-ikx} = 1-(-1) = 2
\end{equation}
which is the difference between the polarization vectors of a zone edge mode on neighboring sites, as required, and gives the maximum potential energy.

The derivative operators act on each component of the polarization vectors ($\epsilon_{\mu}(x)$) which are at each lattice site, i.e. the vertices of the plaquettes, which generates derivatives of both the transverse \textit{and} longitudinal components similarly to the spinor derivative:
\begin{equation}
\partial_{\mu}\epsilon_{\nu}(x)=\frac{1}{|\hat{x}_{\mu}|}\big(\epsilon_{\nu}(x)-e^{ik^{\mu}\hat{x}_{\mu}}\epsilon_{\nu}(x)\big)
\end{equation}
For the longitudinal mode only the derivatives equivalent to $k_{\mu}$ where $k^{\mu}\epsilon_{\mu} \neq0$ will contribute, while for the transverse modes only the components where $k^{\mu}\epsilon_{\mu} = 0$ will be non-zero.   

Thus summing over all plaquettes in 4-dimensional space ($t,x,y,z$) gives the contributions to the kinetic energy (time derivative) and potential energy (spatial derivatives) of the modes in the usual manner. To avoid double counting the derivatives the plaquettes are summed in a checkerboard fashion (see Supporting Information), and a pre-factor of $\frac{1}{3}$ is needed as each derivative is included in three $x^{i}x^{j}$ planes. For example the $x$-derivative occurs in the $xt, xy$ and $xz$ planes. 

As explained above, computationally the approach would be to just compute each time and spatial derivative of the polarization vectors of each occupied mode at each lattice site; the plaquette discussion here is used as it very intuitively gives a picture of how symmetry-breaking arises.

The U(1) bosonic sector of the action is then:
\begin{equation}
S_{U(1)} = \frac{1}{6}\sum_{ijkl}O_{ijkl}
\end{equation}
where the $i,j,k,l$ indices run over the \textit{neighbouring} vertices in the form of plaquettes in the coordinate planes.

\subsection{Electron-Electron interactions and Yang-Mills Theory}
\begin{figure}
\includegraphics[width=\columnwidth]{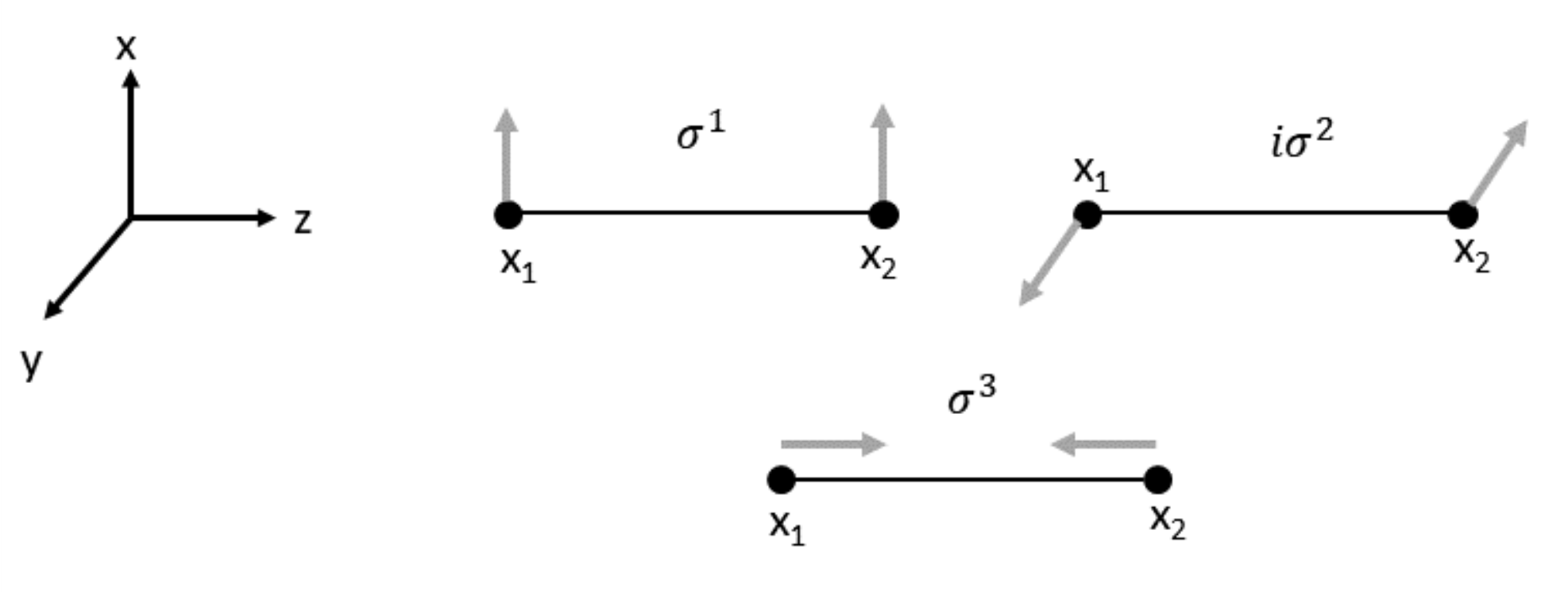}
\caption{\raggedright{Atomic motions which correspond to the different SU(2) modes, $\hat{W}^{1}$, $\hat{W}^{2}$ and $\hat{W}^{3}$.}}
\label{SU2}
\end{figure}

In a previous study\cite{Booth2020} it was demonstrated that the electron-phonon interactions of a system such as VO$_{2}$ which manifests spin and charge ordering during a crystal structure transformation can be described by an interaction vertex which corresponds to an SU(2) Yang-Mills theory. The structural phase transition is characterised by the atomic motions (polarization vectors) of Figure \ref{M2}a, which are a result of the strong electron-electron interactions. Figure \ref{SU2} illustrates the atomic motions associated with the different SU(2) modes.

It is not difficult to see how such polarisation vectors can lower the energy in a strongly correlated system governed by the Hubbard Hamiltonian:\cite{Hubbard1963}
\begin{equation}
H=-t\sum\limits_{\langle ij\rangle}^{}(c^{\dagger}_{i\sigma}c_{j\sigma}+c^{\dagger}_{j\sigma}c_{i\sigma})+\\U\sum\limits_{i}n_{i\uparrow}n_{i\downarrow}
\label{HubbardH}
\end{equation}
where $t$ is the hopping energy, which is proportional to the overlap between neighbouring spinor wavefunctions and $U$ is the on-site energy.

\begin{figure}
\includegraphics[width=0.8\columnwidth]{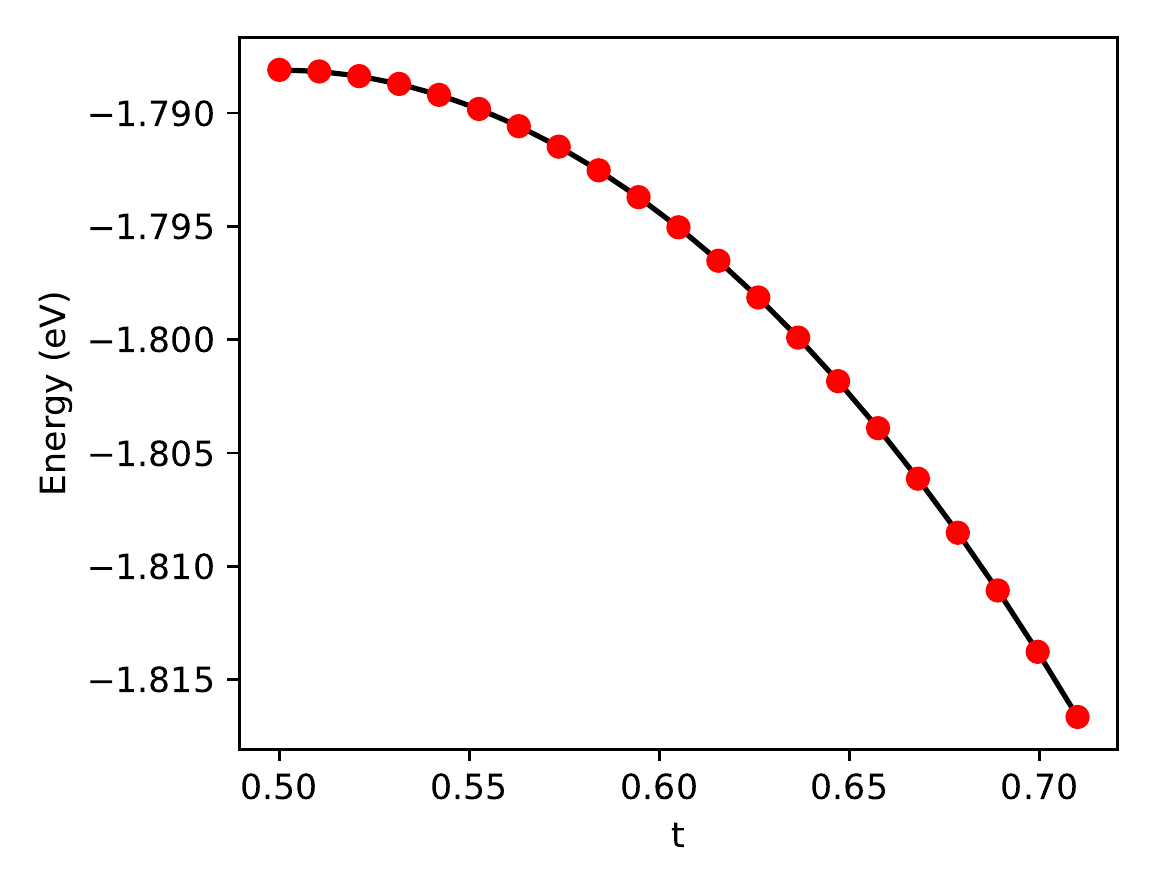}
\caption{\raggedright{Plot of the ground state energy of a six-site 1D Hubbard model with periodic boundary conditions in which the sites pair up, giving a long distance hopping energy of $1-t$ eV and a short distance energy of $t$ eV, with $U = 5$ eV.}}
\label{H_H}
\end{figure}

Figure \ref{H_H} plots the ground state energy of a linear six site system with periodic boundary conditions in which the hopping energies alternate between $t$ and $1-t$, and $t$ increases from 0.5 to 0.75 eV, and $U$ is constant at 5 eV. The $t$ values thus simulate a system in which the atoms are paired, forming short-long-short-...etc inter-atomic distances. 

The data indicates that if the elastic cost of pairing the lattice sites is smaller than the reduction in energy caused by the pairing, then an instability towards a paired system will occur. This is seen in the polymorphs of vanadium dioxide, such as the M$_{1}$ and M$_{2}$ forms of VO$_{2}$, which both exhibit crystal structure transformations characterised by atomic pairing.\cite{Zylberstein1975} 

However, since the kinetic terms ($t$) are proportional to the overlap of the spinor wavefunctions on neighbouring sites (we only include nearest neighbour hopping) this will clearly be a maximum if the spins are antiferromagnetically aligned. In the M$_{2}$ system it was noted that antiferromagnetic alignment coincided with antiferroelectricity,\cite{Pouget1974} which was orthogonal to the pairing along the vanadium atom chains. We proposed that the antiferroelectricity manifests a Rashba-type spin ordering mechanism involving electron-hopping to neighbouring sites.\cite{Booth2020}

The key point here is that what the data of Figure \ref{H_H} suggests is that there is \textit{another source} of phonons in the system, which is the electron correlations. Here we make the assumption that these ``Hubbard" phonons can be decomposed into normal modes in the same manner as the U(1) phonons.

The use of the SU(2) generators to describe the electron-phonon interactions can be most clearly understood by re-imagining the generators of the SU(2) group, the Pauli matrices, to be 2$\times$2 linear transformations acting on neighbouring nuclei. Using $\mathbf{r}_{1}$ and $\mathbf{r}_{2}$ to denote the position vectors $x_{1}$ and $x_{2}$ to avoid ambiguity, schematically we have:

\begin{equation}
	\quad\begin{pmatrix}\hat{z}&\hat{x},\hat{y}\\\hat{x},\hat{y}&\hat{z}\end{pmatrix}\begin{pmatrix}\mathbf{r}_{1}\\\mathbf{r}_{2}\end{pmatrix}
\end{equation}
where hat denotes a vector in the corresponding direction.

For example, schematically the action of the $\sigma^{3}$ matrix is:
\begin{equation}
	\sigma^{3}=\begin{pmatrix}1\hat{z}&0\\0&-1\hat{z}\end{pmatrix}\begin{pmatrix}\mathbf{r}_{1}\\\mathbf{r}_{2}\end{pmatrix}
\end{equation}

Thus it implements a polarization vector in the positive $z$-direction on the nucleus at $\mathbf{r}_{1}$ and at the same time a negative polarization vector along the $z$-axis at $\mathbf{r}_{2}$. Thus this vertex now contains a \textit{pairing} of neighbouring atoms, as per the Peierls distortion of the displacive transitions of VO$_{2}$ (see Figure \ref{SU2}). This pairing mode is given the symbol $\hat{W}^{3}_{\mu}$. This is contracted with the gamma matrices\cite{Booth2020} in the usual manner, however as it points along the $z$-direction it is contracted with $\gamma^{3}$, and doesn't affect the spin orientation, and thus is considered \textit{neutral}.

The $\sigma^{1}$ and $\sigma^{2}$ components generate spin raising and lowering operators \textit{and} antiferroelectricity when contracted with the appropriate gamma matrices as described below, and therefore are considered \textit{charged}. These are given the symbols $\hat{W}^{1}_{\mu}-i\hat{W}^{2}_{\mu}=\hat{W}^{-}_{\mu}$, which acts on position state $x_{1}$, and $\hat{W}^{1}_{\mu}+i\hat{W}^{2}_{\mu}=\hat{W}^{+}_{\mu}$ which acts on position state $x_{2}$. Therefore this vertex can also order neighbouring spins antiferromagnetically, again, reminiscent of the ordering seen in the M$_{2}$ VO$_{2}$ structure.\cite{Pouget1976}

The full electron phonon interaction vertex has the form:
\begin{multline}
	g_{a}\bar{\psi}\gamma^{\mu}\hat{W}^{a}_{\mu}\psi=\\g_{(1,2,3)}\begin{pmatrix}\bar{\psi}^{\prime}_{\mathbf{ab}}(x^{\prime}),\bar{\psi}^{\prime}_{\mathbf{ab}}(x^{\prime})\end{pmatrix}\times\\\gamma^{\mu}\begin{pmatrix}W^{3}_{\mu}(x_{1})&W^{1}_{\mu}+iW^{2}_{\mu}(x_{2})\\W^{1}_{\mu}-iW^{2}_{\mu}(x_{1})&-W^{3}_{\mu}(x_{2})\end{pmatrix}\begin{pmatrix}\psi_{\mathbf{a}}(x_{1}) \\ \psi_{\mathbf{b}}(x_{2})\end{pmatrix}
	\label{e-ph}
\end{multline}
where $g_{(1,2,3)}$ are the couplings to the different modes, the $\psi_{\mathbf{a},\mathbf{b}}(x)$ are 4-component Nambu spinors\cite{Booth2020}, the $x^{\prime}$ coordinates are linear combinations of $x_{1}$ and $x_{2}$ and the gamma matrices are the chiral forms used above:
\begin{equation}
\gamma^{0} = \begin{pmatrix}0&\mathbb{1}\\\mathbb{1}&0\end{pmatrix}, \gamma^{i} = \begin{pmatrix}0&\sigma^{i}\\-\sigma^{i}&0\end{pmatrix}
\end{equation}
and thus
\begin{equation}
\bar{\psi} = \psi^{\dagger}\gamma^{0} = (\bar{\psi}_{\mathbf{a}},\bar{\psi}_{\mathbf{b}}) = (\psi^{\dagger}_{\mathbf{a}}\gamma^{0},\psi^{\dagger}_{\mathbf{b}}\gamma^{0})
\end{equation}

Setting $g_{1}W^{1}_{1} = g_{2}W^{2}_{2} = 1$ and all other polarization vector components to zero we get a term:
\begin{multline}
	\begin{pmatrix}\bar{\psi}_{\mathbf{a^{\prime}}},\bar{\psi}_{\mathbf{b}^{\prime}}\end{pmatrix}\\\times\begin{pmatrix}0&g_{+}(\gamma^{1}+i\gamma^{2})\\g_{-}(\gamma^{1}-i\gamma^{2})&0\end{pmatrix}\begin{pmatrix}\psi_{\mathbf{a}}\\\psi_{\mathbf{b}}\end{pmatrix}
\end{multline}
Remembering that:
\begin{equation}
\gamma^{i} = \begin{pmatrix}0&\sigma^{i}\\-\sigma^{i}&0\end{pmatrix}
\end{equation}
this gives the familiar spin raising and lowering operators, $S^{+}=\sigma^{1}+i\sigma^{2}$, and $S^{-}=\sigma^{1}-i\sigma^{2}$:
\begin{multline}
	\bar{\psi}_{\mathbf{a^{\prime}}}g_{+}\begin{pmatrix}0&\hat{S}^{+}\\-\hat{S}^{+}&0\end{pmatrix}\psi_{\mathbf{b}}(x_{2})\\+\bar{\psi}_{\mathbf{b^{\prime}}}g_{-}\begin{pmatrix}0&\hat{S}^{-}\\-\hat{S}^{-}&0\end{pmatrix}\psi_{\mathbf{a}}(x_{1})
	\label{Spin_Operators}
\end{multline}
with the negative sign in the $\gamma^{i}$ accounting for the opposite helicities of the two-component spinors in each four-component spinor such that the Weyl equation for each is satisfied. This justifies the relabeling of the linear combinations of the $\hat{W}^{1}_{\mu},\hat{W}^{2}_{\mu}$ bosons as $\hat{W}^{+}_{\mu}(x_{1})$ and $\hat{W}^{-}_{\mu}(x_{2})$.

\begin{figure}
\includegraphics[width=\columnwidth]{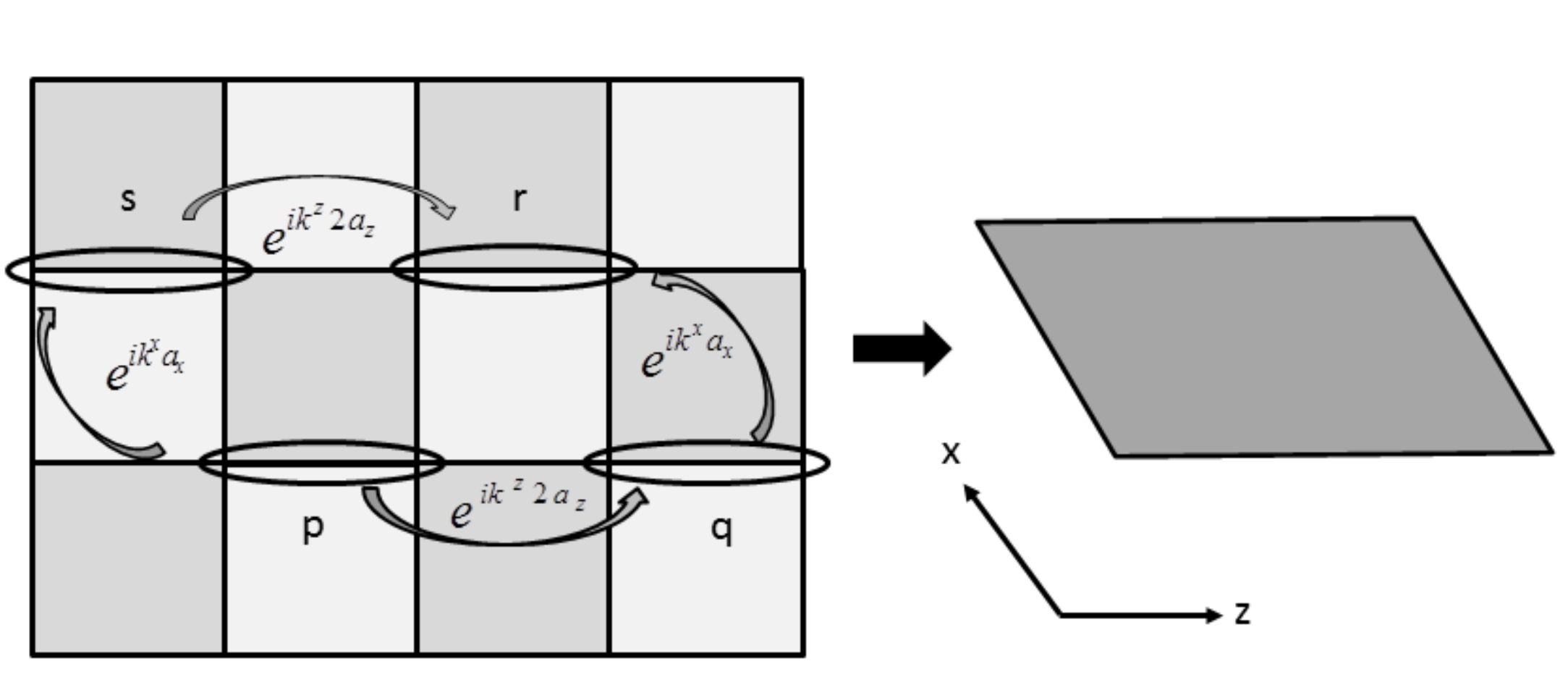}
\caption{\raggedright{Schematic illustration of the plaquette action of a system in which SU(2) Yang-Mills bosons are active. The bosons are off-set vertically in the same manner as the pairing modes in VO$_{2}$, thus the coordinate system of the Yang-Mills fluctuations is different to the high symmetry state.}}
\label{Sym_Break}
\end{figure}

Given that the bosons are operators acting on the position states of the atoms, there is a slight subtlety to the effect of the $\hat{W}^{+}/\hat{W}^{-}$ operators. Since these are composite bosons their actions are defined to be of the form:
\begin{equation}
\hat{W}^{-}|\psi_{\mathbf{a}}(x_{1})\rangle = (\hat{W}^{1}-i\hat{W}^{2})|\psi_{\mathbf{a}}(x_{1})\rangle \sim \hat{a}^{\dagger}\epsilon_{x}\hat{a}^{\dagger}\epsilon_{y}|\psi_{\mathbf{a}}(x_{1})\rangle
\end{equation}
\begin{equation}
\hat{W}^{+}|\psi_{\mathbf{b}}(x_{2})\rangle = (\hat{W}^{1}+i\hat{W}^{2})|\psi_{\mathbf{b}}(x_{2})\rangle \sim -(\hat{a}^{\dagger}\epsilon_{x}\hat{a}^{\dagger}\epsilon_{y})|\psi_{\mathbf{b}}(x_{2})\rangle
\end{equation}
where the imaginary unit and details of the contraction with the gamma matrices are omitted for clarity. Thus although only the $\sigma_{2}$ generator has a minus sign, the operator products for $\hat{W}^{+}/\hat{W}^{-}$ produce polarization vectors in opposite directions, which provides the required antiferroelectricity.

Applying the full vertex (i.e. non-zero polarization vector components for $\hat{W}^{3}_{\mu}$ as well as the other two) along chains of atoms with for example a single itinerant electron inhabiting a sum of tight-binding momentum states at each site which are themselves in some parallel arrangement giving a three-dimensional crystal (similarly to Tetragonal VO$_{2}$), the temporal and spatial variation of each SU(2) mode is given by:
\begin{equation}
e^{-ip_{\mu}x^{\mu}} = e^{i(\mathbf{k}\mathbf{x}-\omega t)}
\end{equation}
where it is assumed that the SU(2) mode is collective, i.e. the polarization vectors are applied to each of the two sites and have the same magnitude, and the vectors on the two sites vary in time and space with the phase above.

For these collective modes we have the situation illustrated in Figure \ref{Sym_Break}. The electron-electron interactions now result in the pairing and spin-ordering modes of the SU(2) theory, however above T$_{c}$ there is still enough disorder in the system that they are not yet static nor are they in phase with each other. We then have the same situation as for the individual vector bosons, the phases describe the differences between the polarisation vectors at each pair of sites.

Since the the modes are constructed by dotting polarisation vectors into the Pauli matrices, for example:
\begin{equation}
\hat{W}^{3}(x_{1},x_{2}) = \epsilon_{z}\sigma^{z} = \begin{pmatrix}\epsilon_{z}(x_{1})&0\\0&-\epsilon_{z}(x_{2})\end{pmatrix}
\end{equation} 
the variation of the polarisation vector is given as per a U(1) mode, however with a lattice vector now double the length in the $x_{2}-x_{1}$ direction, but they remain the same in each other direction. So in the case of $x_{3},x_{4}$ being collinear along the $\hat{z}$-direction and adjacent to $x_{1},x_{2}$:
\begin{equation}
\hat{W}^{3}(x_{3},x_{4}) = e^{ik^{z}2\hat{x}_{z}}\begin{pmatrix}\epsilon_{z}(x_{1})&0\\0&-\epsilon_{z}(x_{2})\end{pmatrix}
\end{equation}
The phases in the other directions remain the same as the U(1) theory. However, in Figure \ref{Sym_Break} the bosons are off-set in the vertical direction, similarly to the case of VO$_{2}$, which is assumed to be for reasons of structural rigidity. Thus the coordinate system which describes the changes of the Yang-Mills bosons is different to that of the high symmetry state. 

The full boson action for the system then becomes:
\begin{equation}
S = S_{U(1)} + S_{SU(2)} = \frac{1}{6}\bigg(\sum_{ijkl}O_{ijkl} + \sum_{pqrs}O_{pqrs}\bigg)
\label{Full_Action}
\end{equation}
where $i,j,k,l$ are the vertices of the U(1) plaquettes, and those of the SU(2) plaquettes are $p,q,r,s$ and the plaquettes are functions of space and time.

The key element of this action is that by including the SU(2) bosons, which are based on the data of Figure \ref{H_H} and experimental observations,\cite{Goodenough1971,Zylberstein1975} the electron-electron interactions are being included by proxy. Thus a Hamiltonian derived from this action will provide a way of minimising the potential energy of the electron liquid.

\subsection{Phase Coherence and Critical Behaviour}
\begin{figure}
\includegraphics[width=0.9\columnwidth]{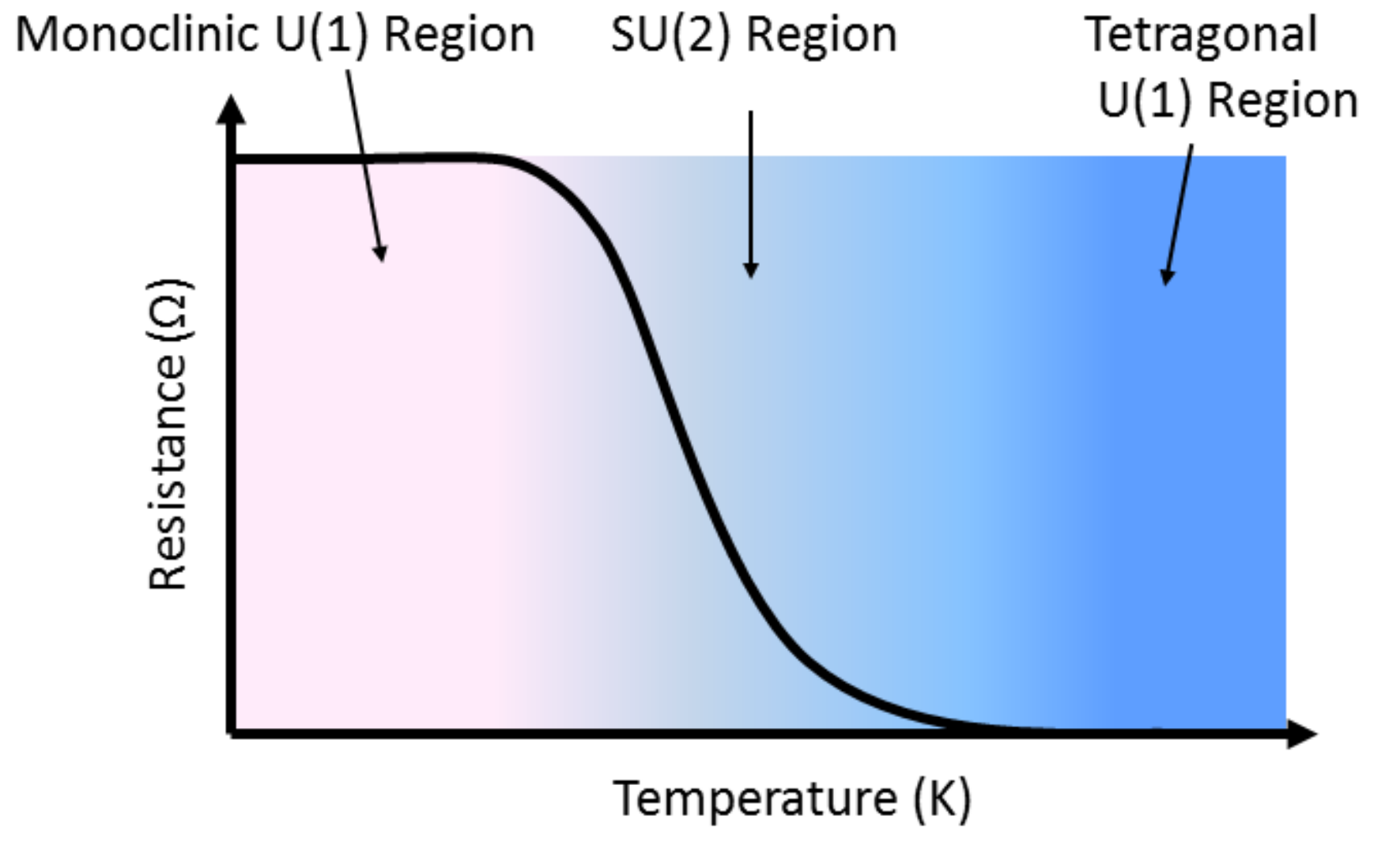}
\caption{\raggedright{Schematic illustration of the Resistance versus Temperature behaviour of VO$_{2}$ which indicates the transition from the high symmetry tetragonal metallic phase to the low symmetry insulating monoclinic phase. At high Temperature the phonons will correspond to a standard U(1) vector field, however as the Temperature decreases, strong electron correlations result in SU(2) bosons dominating the lattice fluctuations near the transition.}}
\label{Phases}
\end{figure}

Figure \ref{Phases} presents a schematic illustration of the Resistance versus Temperature behaviour of a system undergoing a metal insulator transition. In this example, vanadium dioxide, the system transitions from a high-symmetry tetragonal metallic state to a low symmetry monoclinic insulating state.

At $T >> T_{c}$ the U(1) term of the action is expected to dominate as the amount of energy in the lattice is so great that the fluctuations caused by the electron-electron interactions are ``washed out". It might be expected though, that as $T \rightarrow T_{c}$ the temperature becomes low enough that the polarization vectors created by the electron-electron interactions \textit{via} the SU(2) electron-phonon interaction vertex are able to compete with the polarization vectors of the U(1) action. Thus as we approach T$_{c}$ from above the SU(2) sector of the action will increasingly dominate the system's behaviour.

The energy-wavevector dependence of the U(1) modes is derived from the relationship between the kinetic and potential energies of atoms interacting \textit{via} a harmonic potential. However the relationship will be different for the SU(2) modes, as the polarisation vectors are a result of electron-electron interactions. Taking the $\hat{W}^{3}$ boson as an example; obviously, if the phases differ from site-to-site, the the system is not paired in the manner that is seen in the ground state of the VO$_{2}$ system, or the data of Figure \ref{H_H}. 

However, if the bosons are all in phase and the $\hat{W}^{1}$ and $\hat{W}^{2}$ bosons have ordered the spins antiferromagnetically, then this describes perfect ordering of the pairs of atoms, which lower the energy as per Figure \ref{H_H}. This situation corresponds to the $\mathbf{k} = 0$ ($\omega$ may still be finite) case of Figure \ref{Sym_Break}, that is the bosons are \textit{phase coherent}. However, if the energy of the mode ($\omega$) is not zero, the time derivative of the mode will not be zero, and this describes a system in which the atoms oscillate between a Short-Long-Short... configuration and a Long-Short-Long... configuration. 

This is obviously higher energy than the paired state as it passes through the symmetric structure, and thus the energy of the electron component of the Hamiltonian will oscillate between the maximum and minimum vales of Figure \ref{H_H}. Thus, the lowest energy mode of the Yang-Mills system will also have zero time-derivative, that is it will ``soften": $\omega \rightarrow 0$, and the polarization vectors will ``freeze" into the structure.

This can be conveniently expressed in the Hamiltonian:
\begin{equation}
\hat{H} = -J\sum_{a,x, \hat{x}} \textrm{tr}\big(\hat{W}^{a}(x,t)\hat{W}^{a}(x+\hat{x},t)\big)
\end{equation}
where $a$ labels the Yang-Mills boson, and $\{\hat{x}\}$ is a set of vectors which describes the arrangements of the Yang-Mills fluctuations in space, for example in Figure \ref{Sym_Break} these describe the parallelogram structure. The parameter $J$ is phenomenological and describes the coupling of the Yang-Mills bosons, which is expected to increase as the Temperature is lowered (electron-electron interactions become increasingly important), and is thus the quantity $-J$ is an effective Temperature.

\begin{figure}
\subfigure{\includegraphics[width=0.9\columnwidth]{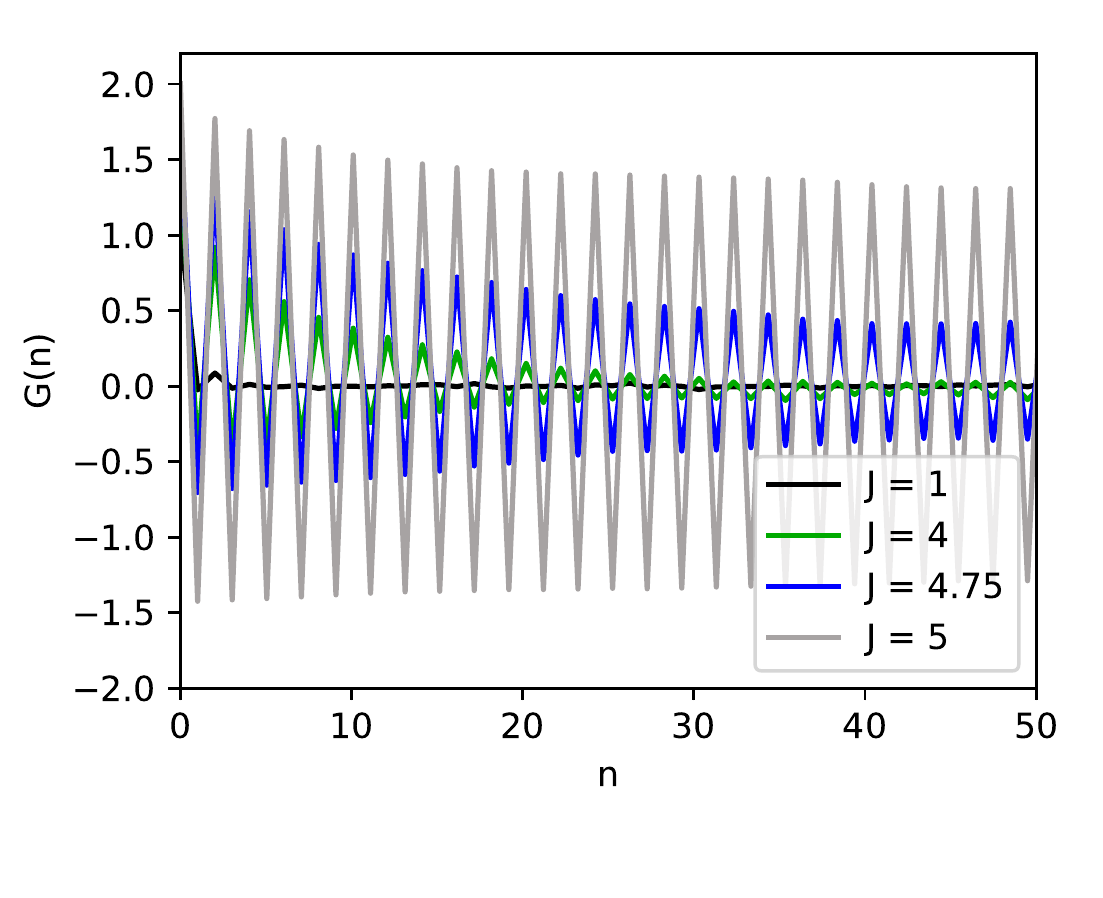}}
\subfigure{\includegraphics[width=0.85\columnwidth]{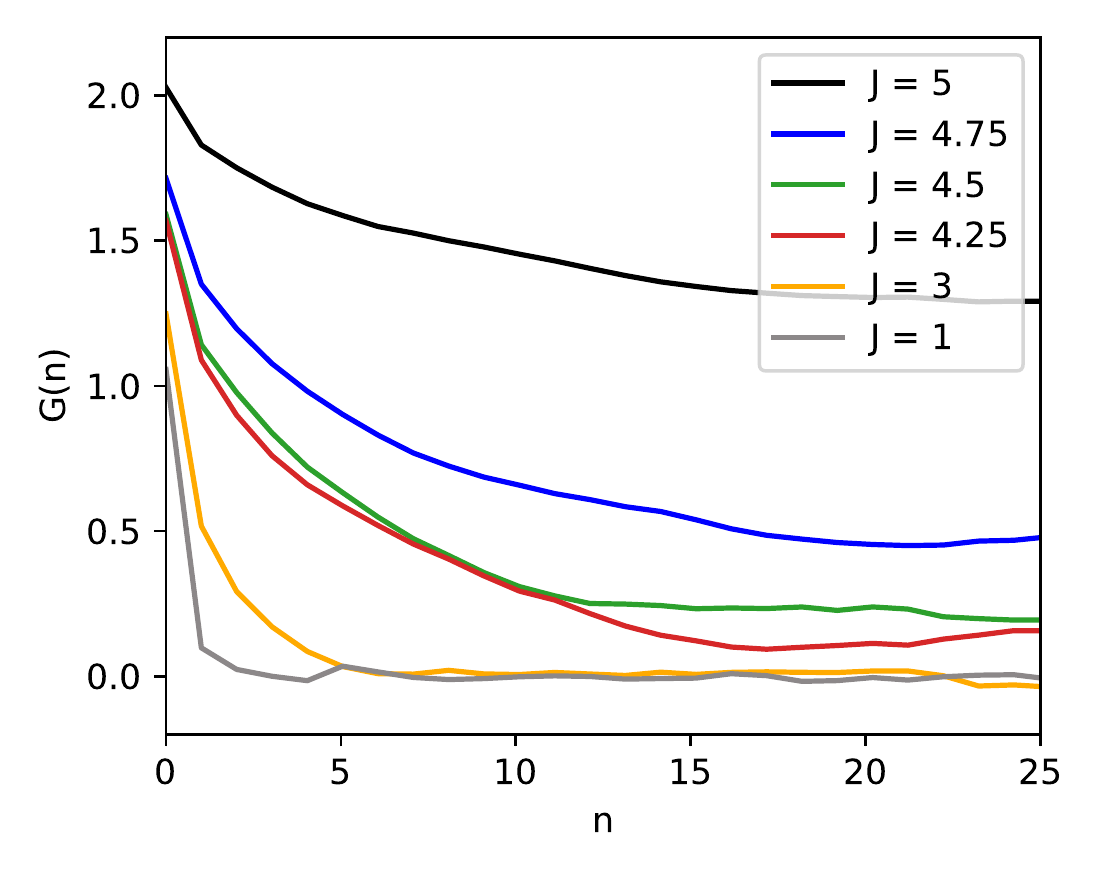}}
\subfigure{\includegraphics[width=0.9\columnwidth]{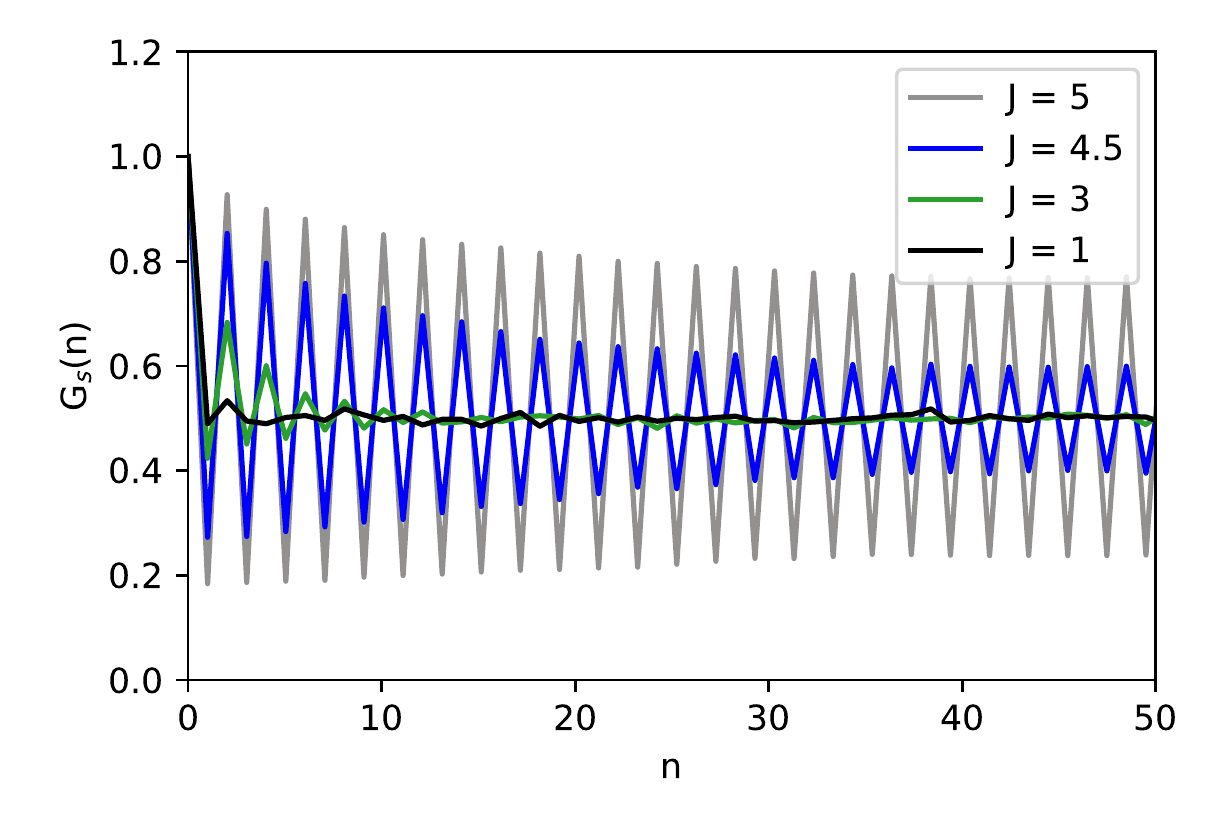}}
\caption{\raggedright{Monte Carlo correlation functions on a 100-site chain of atoms, of a) the polarization vectors at each site showing increasing anticorrelation as the coupling (or effective Temperature) increases, and b) the polarization vectors of the Yang-Mills modes showing increasing correlation as the coupling (effective Temperature) increases, and c) the spin-spin correlation function showing increasing antiferromagnetism as coupling increases.}}
\label{Corr_Fns}
\end{figure}

Figure \ref{Corr_Fns} presents Monte Carlo calculations of time-averaged correlation functions of the U(1) phonon polarization vectors in Figure \ref{Corr_Fns}a and the Yang-Mills bosons in Figure \ref{Corr_Fns}b, and the spin-spin correlation function in Figure \ref{Corr_Fns}c, computed on a chain of 100 sites with periodic boundary conditions.

The calculation starts from a random configuration of (classical) spins and polarization vectors, and time evolves according to the boson action (Equation \ref{Full_Action}) using a Metropolis algorithm. 

The data of Figure \ref{Corr_Fns}a illustrates that as the effective Temperature ($J$) decreases (increasing coupling) the polarization vectors of the atoms become increasingly anti-correlated, indicating that the nuclei are pairing. Another perspective of this is presented in Figure \ref{Corr_Fns}b, which plots the correlation function of the Yang-Mills bosons, indicating that they become increasingly correlated as the Temperature decreases.

Figure \ref{Corr_Fns}c plots the spin-spin correlation function as a function of effective Temperature, and clearly illustrates that like the U(1) polarization vectors, the spins become increasingly anti-correlated as Temperature decreases.

\begin{figure}
	\includegraphics[width=0.75\columnwidth]{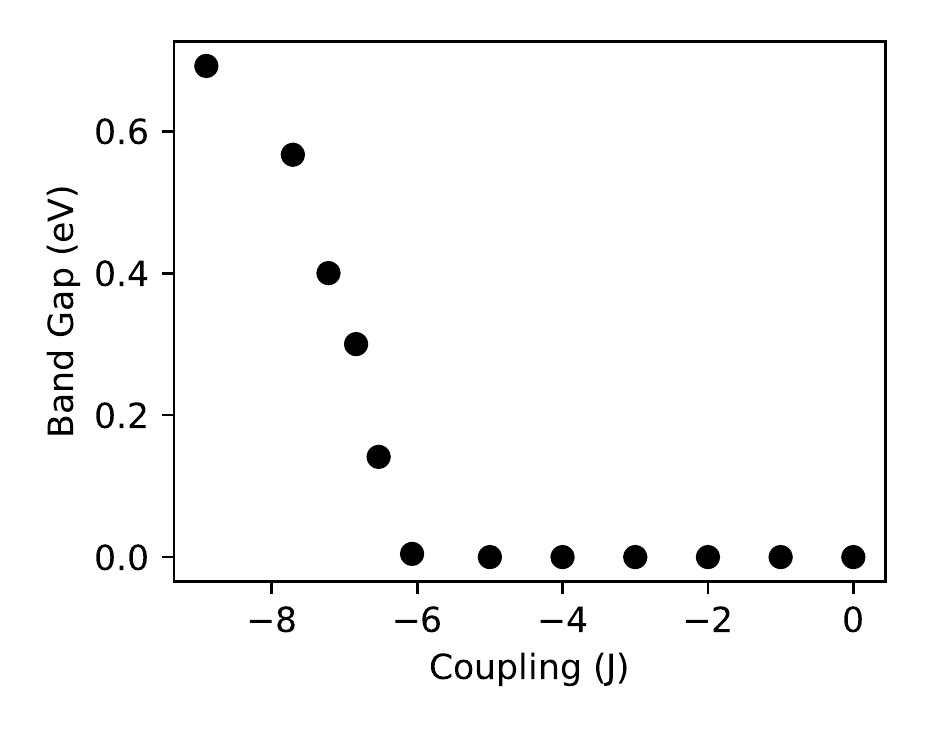}
	\caption{\raggedright{A plot of GW Band Gaps of vanadium dioxide structures whose time-averaged atomic positions correspond to different values of the effective Temperature $J$.}}
	\label{Band_Gaps}
\end{figure}

Figure \ref{Band_Gaps} presents band gaps of VO$_{2}$ structures plotted against the corresponding value of the effective Temperature $-J$. The data exhibits an opening of the VO$_{2}$ band gap at $J \sim -6$.

Altogether, the data characterises a system in which the time averages of the electron spins and position fluctuations of the nuclei describe a system in which the atoms pair up into a ``Peierls-Paired"\cite{Goodenough1971} configuration, while at the same time antiferromagnetically ordering the electron spins.

If one imagines the system in Figure \ref{Sym_Break} to consist of the same structure layered in the $\hat{y}$-direction, the high symmetry cubic structure will break its symmetry and form a monoclinic structure. In this case this is because of the off-set order of the SU(2) interaction vertices which are directed down the $\hat{z}$-direction, thus the planes arranged in the $\hat{y}$-direction will be identical. 

Given the combination of pairing from the $\hat{W}^{3}$ mode and the spin ordering from the $\hat{W}^{1}$ and $\hat{W}^{2}$ modes, the electron-phonon interaction vertex of equation \ref{e-ph}, being a $2\times 2$ linear transformation, and the data of Figure \ref{Corr_Fns} predict that static linear combinations of electron wavefunctions will result, and the system will exhibit a static charge density wave order: 
\begin{multline}
\sum_{a}\hat{W}^{a}\begin{pmatrix}
\psi_{\mathbf{a}}(x_{1})\\\psi_{\mathbf{b}}(x_{2})\end{pmatrix} = \\ \begin{pmatrix}\langle\hat{W}^{3}(x_{1})\rangle\psi_{\mathbf{a}}(x_{1}) + \langle\hat{W}^{+}(x_{2})\rangle\psi_{\mathbf{b}}(x_{2})\\ \langle\hat{W}^{-}(x_{1})\rangle\psi_{\mathbf{a}}(x_{1}) - \langle\hat{W}^{3}(x_{2})\rangle\psi_{\mathbf{b}}(x_{2})\end{pmatrix}
\end{multline}
where the $\langle\dots\rangle$ terms indicate the vacuum expectation values of the polarization vectors. 

Therefore we have linear combinations of the original position state wavefunctions, corresponding to bonding and anti-bonding states. However these linear combinations have the symmetry of the SU(2) bosons, and therefore in this state construction of the electron momentum states will involve taking these linear combinations using the phases and coordinate system of the Yang-Mills description in Figure \ref{Sym_Break}.

Constructing electron momentum states now proceeds by using these linear combinations, e.g.:
\begin{equation}
\psi_{\mathbf{k}}(\mathbf{r})=\sum_{\mathbf{R}}\bigg(\langle\hat{W}^{3}\rangle\psi_{\mathbf{a}}(\mathbf{r}-\mathbf{R}) + \langle\hat{W}^{+}\rangle\psi_{\mathbf{b}}(\mathbf{r}-\mathbf{R})\bigg)e^{i\mathbf{kR}}
\end{equation}
where now $\mathbf{a}$, $\mathbf{b}$ label the positions of the basis wavefunctions: the original single particle position states before symmetry-breaking. The lattice fluctuations now correspond to constructing a U(1) theory in the same manner as for the Tetragonal phase, but with the vectors $\mathbf{R}$ corresponding to the new symmetry-broken state. This describes the acoustic modes of the symmetry-broken state and corresponds to the Monoclinic U(1) region labelled in Figure \ref{Phases}.

At $T >> T_{c}$ the optical U(1) fluctuations of the Monoclinic structure will disturb the paired atoms, and the attractive interaction between them will disappear, which is the analogue of asymptotic freedom, represented by the $J = 1$ curves of Figure \ref{Corr_Fns}, in which charge and spin become de-confined. 

Thus this formalism indicates that strong-electron correlations in systems such as vanadium dioxide can indeed result in symmetry-breaking crystal structure transformations which manifest both spin and charge order.

We see that within this formalism it is easy to describe symmetry-breaking transformations which are characterised by both lattice changes, and spin- and electron charge ordering. It will be of significant interest to apply this formalism to doped systems, and determine the characteristics of the electron liquid when both the high- and low symmetry bosonic fluctuations are active.

\section{Conclusions}
Reformulating the physics of tight-binding electron momentum states and phonon modes in a Wilsonian-like fashion allows a simple action to be derived which describes crystal systems in a manner which is easily adapted to symmetry-breaking. Applying this approach using an SU(2) formalism derived for the case of a strongly-correlated system analogous to vanadium dioxide in which neighboring atoms pair up to lower the energy allows a simple description of symmetry-breaking phase transitions to be developed which includes strong-electron correlations intrinsically. 

Monte Carlo calculations implementing time-evolution using a metropolis algorithm confirm a phase transition occurs as the SU(2) electron-phonon coupling strength increases, which coincides with antiferromagnetic spin ordering. 

This work shows that there is a remarkable similarity between the Weak Interaction sector of the Standard Model of Particle Physics and the \textit{microscopic} mechanism of symmetry-breaking in quasi-linear strongly correlated metal oxides. 

The most natural question to ask then is, are there more similarities to explore, and what can we learn from them? There is at least one more remarkable coincidence between the Standard Model and Materials Science, and that is that the lattice fluctuations of strongly correlated hexagonal systems such as graphene can be described by an SU(3) gauge theory analogously to Quantum Chromodynamics.\cite{Booth_SU3} Thus it may be that systems such as VO$_{2}$ and graphene are not just of enormous significance to materials scientists for new devices, but perhaps also to high energy physicists as new laboratories in which to more conveniently explore Yang-Mills physics.

\section{Methods}
There are two components to the boson action of equation \ref{Full_Action}. The first term describes the ``normal" or thermal phonons of the system, i.e. those that do not result from the electron-electron interactions. At a finite transition Temperature these effectively contribute a background ``noise" which competes with the Yang-Mills bosons (the second term). 

However, there is a slight subtlety to this noise. At the T$_{c}$ of vanadium dioxide, 340 $K$, it is expected that there will be occupation of the optical modes in addition to the acoustic modes. Therefore there will be a slight tendency of the vanadium atoms to move in opposite directions, or for them to be slightly anti-correlated. Correlation would of course correspond to low energy acoustic modes, in which the polarization vectors move in the same directions.

Therefore action used in the Metropolis algorithm\cite{Creutz1983} was based on an Ising-type Hamiltonian:
\begin{multline}
	\hat{H} = -T\sum_{x, \hat{x}}\hat{A}(x,t)\hat{A}(x+\hat{x},t)\\-J\sum_{a,x, \hat{x}} \textrm{tr}\big(\hat{W}^{a}(x,t)\hat{W}^{a}(x+\hat{x},t)\big)
	\label{ActionH}
\end{multline}

where the first term describes the U(1) modes, and the $T$ parameter sets the magnitude of the tendency toward optical excitations. $T$ was set to 0.2 for all simulations, while the Yang-Mills coupling $J$ was varied 1 to 25. This ratio was found to give a transition at a value of the Yang-Mills coupling which is of similar magnitude to the value of the Hubbard $U$ employed in Dynamical Mean Field Theory calculations (4 eV\cite{Biermann2005,Tomczak2007}) and in the exact diagonalization calculations described below (5 eV).

The Monte Carlo calculations of the correlation functions of the polarisation vectors and spins were performed on a chain of 100 sites using a Metropolis algorithm starting with a random configuration of 3-vectors and spins, with the spins restricted to spin-up and spin-down only. For each value of the coupling $J$, the sites were thermalised using 50 sweeps, and then 100 updates were performed for each of 100 correlation functions, which were then averaged.

The correspondence between the coupling $J$ and the transitional VO$_{2}$ structures whose band gaps were calculated using the GW approximation was done using Monte Carlo calculations on a chain of 10 sites with the same action as that for the polarisation vetors. The system was thermalised using 50 sweeps, and the positions averaged 50 times with 50 updates between averages. This procedure was repeated 2,500 times to generate a mean displacement, which was then correlated with the transitional structures generated as described below.

To generate the transitional VO$_{2}$ structures the crystal structure parameters of VO$_{2}$ were obtained from the literature\cite{Andersson1954}. All \textit{ab initio} calculations performed using Projector Augmented Waves\cite{Blochl1994b} and the Vienna Ab Initio Simulation Package (VASP)\cite{Kresse1996}. The DFT functional used in all calculations was the GGA functional of Perdew, Burke and Ernzerhof\cite{Perdew1996}. The literature Monoclinic and Tetragonal structures were relaxed to their respective ground states using Methfessel and Paxton smearing\cite{Methfessel1989} and a conjugate gradient algorithm. 

The atomic positions of a 1$\times$1$\times$2 supercell of the tetragonal structure were then subtracted from those of the monoclinic structure, which generated vectors describing the movement of the atoms across the transition. Vectors describing the changes in unit cell dimensions were obtained in the same manner.  These vectors were then divided such that 10 structures were generated, with the monoclinic structure being the first, and the tetragonal being the last. 

Thus each structure corresponds to a value of the prefactor which multiplies the atomic displacement, and unit cell modifications that occur across the transition. This prefactor was then correlated with the average displacement produced by the Monte Carlo calculations, in this manner the transition structures can be correlated with a value of the Yang-Mills coupling $J$.

The nudged elastic band technique\cite{Henkelman2000} was then applied to these structures, in order to find the minimum energy path between them. Finer resolution was then obtained by applying the same process to 3 structures intermediate to Steps 1 and 2, and Steps 2 and 3.

The GW band structure and input data for the Bloch Equations  were calculated on a 6$\times$6$\times$6 Monkhorst-Pack\cite{Monkhorst1976} k-point mesh using a single-shot G$_{0}$W$_0$ approach\cite{Shishkin2006}. All GW calculations used the Tetrahedron method for Brillouin Zone integration with Bl\"ochl corrections\cite{Bloechl1994}, using a grid of 30 frequency points and 192 bands. 

The energies of the Hubbard Model ground states with different hopping energies in the Hamiltonian were obtained by using the Python package QuSpin.\cite{Weinberg2019}

\section{Acknowledgements}
The author acknowledges valuable support from S. Russo, and useful conversations with S. Todd and S. Bilson-Thompson. Requests for materials should be addressed to jamie.booth@rmit.edu.au
\section{References}
\bibliography{library}

\end{document}